\documentclass[epj]{webofc}

\usepackage[varg]{txfonts}   
\usepackage{hyperref}

\woctitle{XLVI International Symposium on Multiparticle Dynamics}

\begin{document}

\title{On wounded constituents in nuclear collisions\footnote[1]{Presented by M. Rybczy\'{n}ski}}

\author{Piotr Bo\.{z}ek\inst{1}\fnsep\thanks{\email{Piotr.Bozek@fis.agh.edu.pl}} \and
        Wojciech Broniowski\inst{2,3}\fnsep\thanks{\email{Wojciech.Broniowski@ifj.edu.pl}} \and
        Maciej Rybczy\'{n}ski\inst{3}\fnsep\thanks{\email{Maciej.Rybczynski@ujk.edu.pl}}
}

\institute{AGH University of Science and Technology, Faculty of Physics and
Applied Computer Science, \\ 30-059 Cracow, Poland 
\and
The H. Niewodnicza\'{n}ski Institute of Nuclear Physics,
Polish Academy of Sciences, 31-342 Cracow, Poland
\and
Institute of Physics, Jan Kochanowski University, 25-406 Kielce, Poland
          }

\abstract{%
In this talk we summarize the main results of our recent paper~\cite{Bozek:2016kpf}, where we explore predictions of the wounded quark model 
for particle production and the properties of the initial state formed in ultra-relativistic collisions of atomic nuclei.
}

\maketitle

Particle-particle interactions at high energies usually lead to copious production of new particles whose number 
rises as a function of the collision energy. 
The charged particle multiplicity ($N_{\rm ch}$) and its pseudorapidity density ($dN_{\rm ch}/d\eta$) 
are fundamental measurable quantities which serve as important characteristics of the global properties of the system. 
The data on production of particles in relativistic collisions of atomic nuclei collected during the operation of RHIC~\cite{Back:2001xy,Back:2004dy} 
and the LHC~\cite{Aamodt:2010cz} accelerators indicate that the standard wounded nucleon model~\cite{Bialas:1976ed} (according to the definition, 
a wounded nucleon is the one which underwent at least one inelastic collision) does not describe the observed centrality dependence of 
particle multiplicities in the A+A collisions, unless a component proportional to the binary collisions~\cite{Kharzeev:2000ph} is added.

\begin{figure}[h]
\begin{center}
\includegraphics[width=10cm]{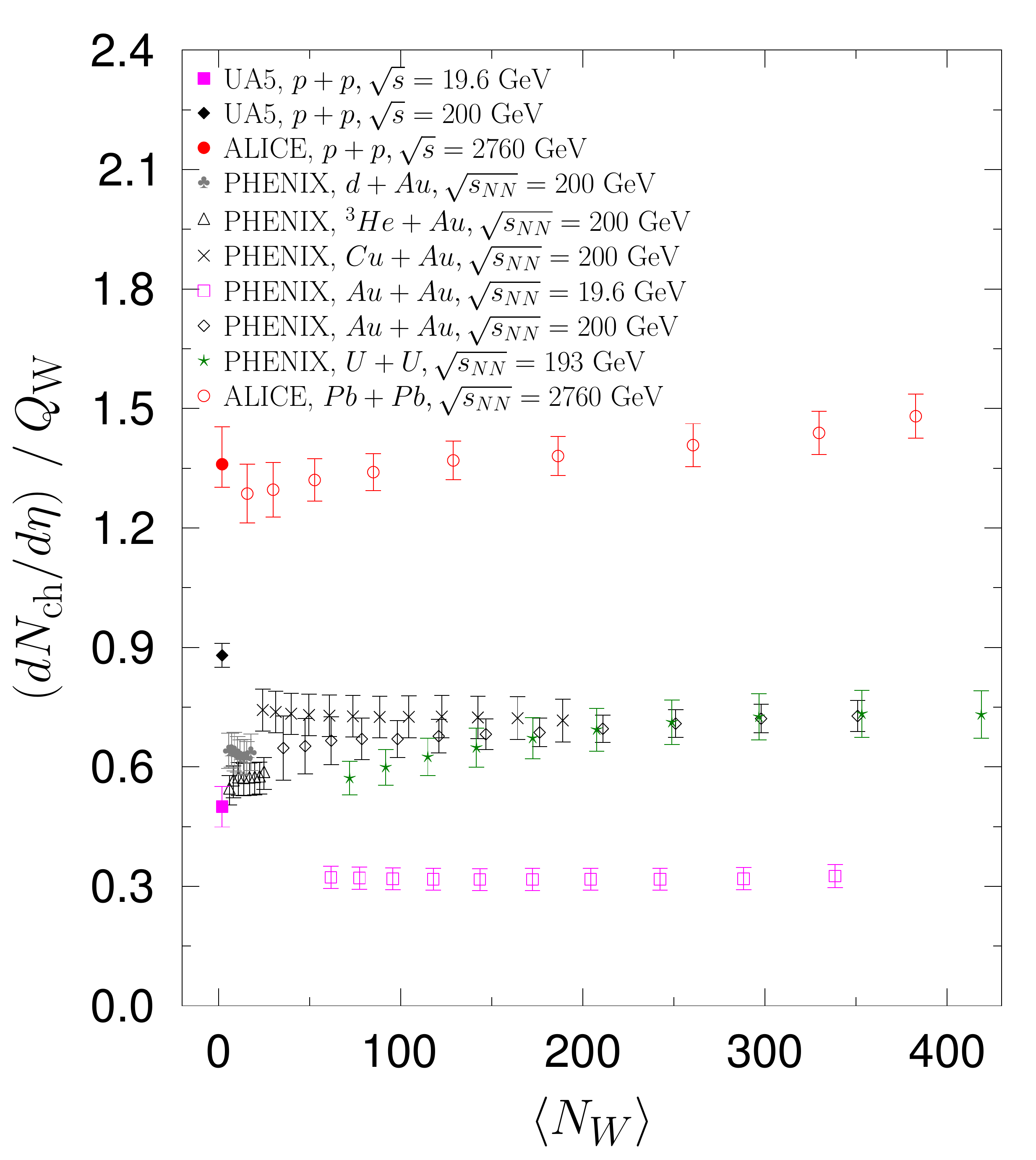}
\caption{\label{fig:dndeta_all} Experimental multiplicity of charged hadrons per
unit of pseudorapidity (at mid-rapidity) divided by the number
of wounded constituents, $dN_{\rm ch}/d\eta/Q_{\rm W}$, plotted as function of
centrality expressed via the number of wounded nucleons.
We also show the results for the p+p collisions (filled symbols
at $\langle N_{\rm W}\rangle = 2$). The data are from
Refs.~\cite{Alner:1986xu,Aamodt:2010cz,Adare:2015bua,Adam:2015gka}.}
\end{center}
\end{figure}

We are interested in the impact of subnucleonic components of matter on the dynamics of the of the early stage of 
the collisions, and explore th wounded quark model~\cite{Bialas:1977en,Bialas:1977xp,Bialas:1978ze,Anisovich:1977av}.
One of the interesting results of~\cite{Bozek:2016kpf} is that taking into account subnucleonic degrees of freedom in the description of the early phase 
of the collision one obtains the linear dependence of production of particles as a function of the number of wounded constituents, namely
\begin{equation}
dN_{\rm ch}/d\eta\propto Q_{\rm W},
\label{eq:dndeta}
\end{equation}
where $Q_{\rm W}$ is the number of wounded constituents involved in the collision. Such scaling has been  suggested in~\cite{Eremin:2003qn} 
for the RHIC data. The wounded quark scaling also works for the SPS energy range~\cite{KumarNetrakanti:2004ym}. 
More recent studies were reported in~\cite{Zheng:2016nxx,Lacey:2016hqy,Mitchell:2016jio,Loizides:2016djv}. 

Our results for particle production
at midrapidity are shown in Fig.~\ref{fig:dndeta_all}. We note an approximately flat centrality dependence of $dN_{\rm ch}/d\eta/Q_{\rm W}$ in qualitative 
agreement with the earlier studies~ \cite{Eremin:2003qn,KumarNetrakanti:2004ym,Adler:2013aqf,Adare:2015bua}. In contrast, the flatness is not the case 
in the wounded nucleon model, where, as is well known, the ratio $dN_{\rm ch}/d\eta/N_{\rm W}$ substantially increases with the number of wounded 
nucleons, $N_{\rm W}$. The value of $dN_{\rm ch}/d\eta/Q_{\rm W}$ (i.e., the average number of charged hadrons per unit of rapidity coming from a 
single wounded quark), is at a given energy roughly similar for various considered reactions. For the LHC energies it is also consistent with the 
production in elementary p+p 
collisions. However, at RHIC energies the p+p point is noticeably higher (about $30\%$) than the band for the A+A collisions.
\begin{figure}[h]
\begin{center}
\includegraphics[width=10cm]{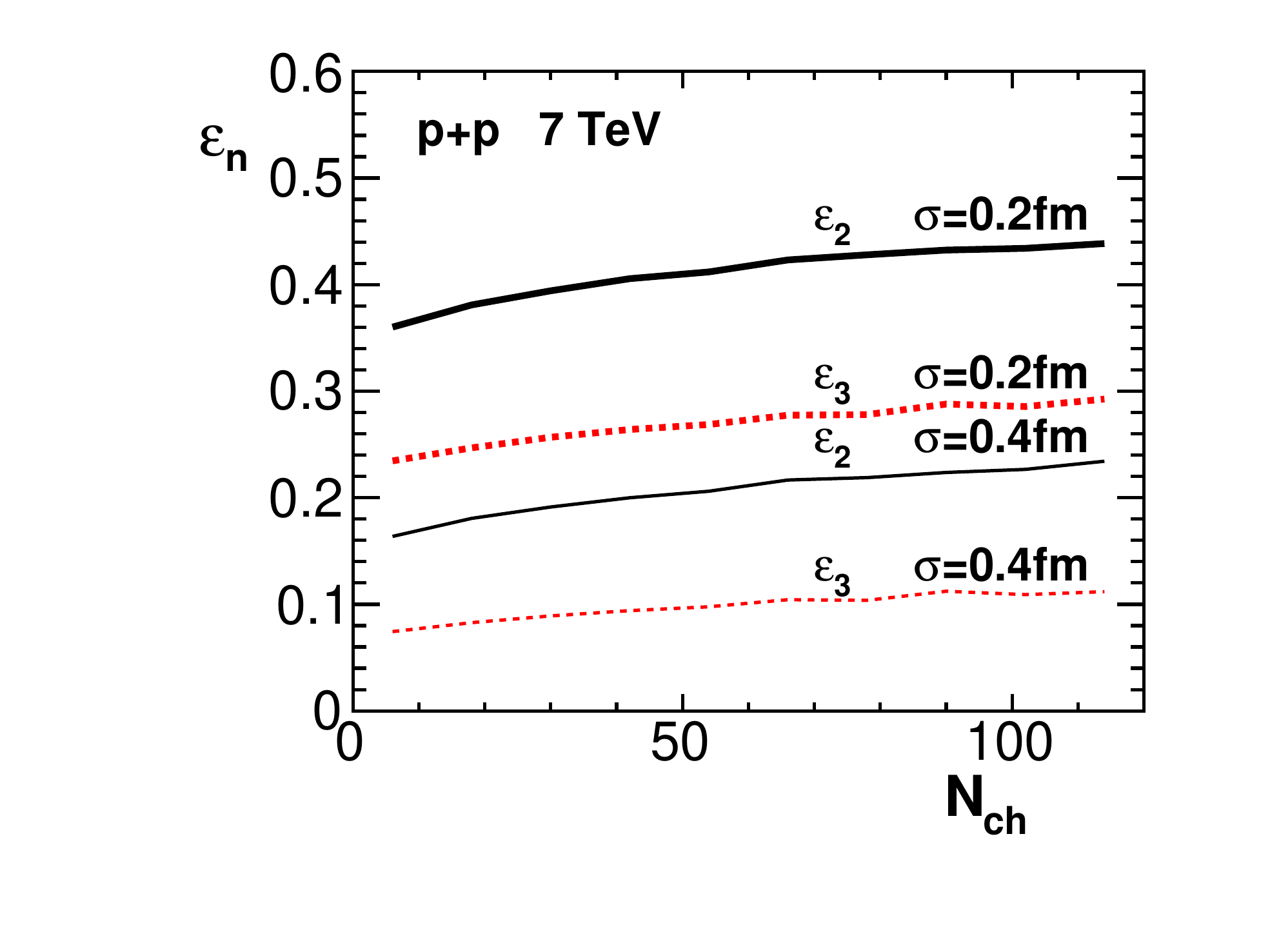}
\caption{\label{fig:ecc} Ellipticity $\varepsilon_{2}$ (solid lines) and triangularity $\varepsilon_{3}$ (dotted lines) of the fireball in 
p+p collisions at $\sqrt{s}=7$~TeV. The eccentricities are plotted as functions of the mean charged multiplicity at $|\eta| < $~2.4.}
\end{center}
\end{figure}

The basic effect of the subnucleonic degrees of freedom in particle production is a stronger combinatorics when compared to 
wounded nucleon model, which accomplishes the approximately linear scaling of production with the number of constituents. 
An intermediate combinatorics is realized in the quark-diquark model~\cite{Bialas:2006qf}, which led to proper description 
of the RHIC data as well as the proton-proton scattering amplitude (including the differential elastic cross section) at the CERN ISR energies.

Modeling with use of subnucleonic components allows for examination of p+p collisions with the methods typically 
used for larger systems. Since on physical grounds the sources of particle production must have a certain transverse 
size, we use Gaussian smearing for the entropy density from a single source centered at a transverse point ($x_{0}$, $y_{0}$), namely
\begin{equation}
g(x,y)=\frac{1}{2\pi \sigma^2} \exp \left ( - \frac{(x-x_0)^2+(y-y_0)^2}{2\sigma^2}\right ).
\end{equation}
It should be noted that in small systems the fireball eccentricities depend largely on the value of the smearing width 
$\sigma$. In Fig.~\ref{fig:ecc} we show fireball eccentricities resulting from the wounded constituents in p+p collisions 
at the LHC. Both the ellipticity and triangularity are large, hence may lead to substantial harmonic flow, in 
accordance to the collectivity mechanism expected for collisions of small systems~\cite{Bozek:2011if,Bozek:2012gr,Bozek:2013uha,Bozek:2013ska} 
at sufficiently high multiplicity. The eccentricities do not depend strongly on the multiplicity of produced particles (centrality), 
which shows that they originate from fluctuations and not the geometry of the collision.

\begin{figure}[h]
\begin{center}
\includegraphics[width=10cm]{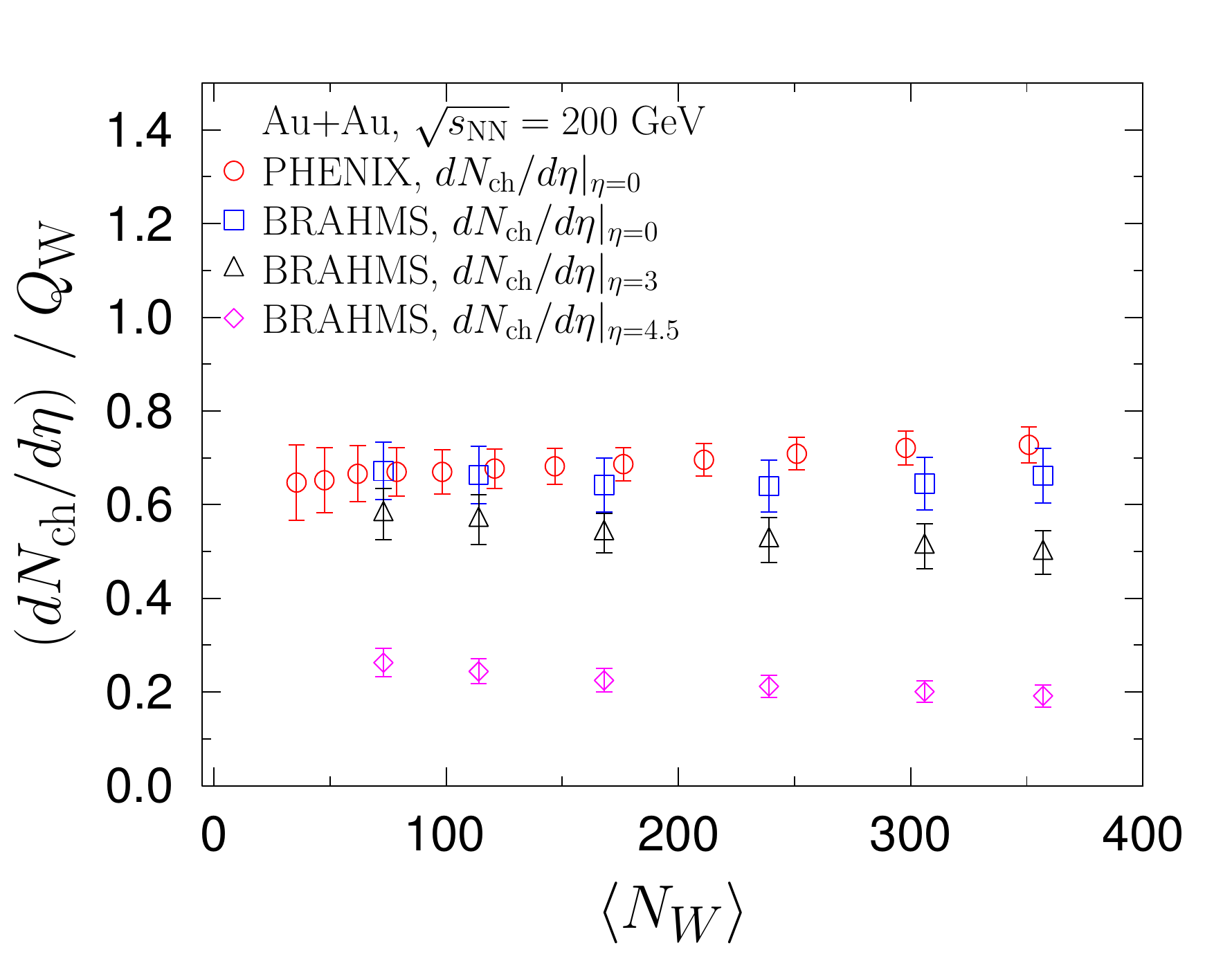}
\caption{\label{fig:dndeta} Experimental multiplicity of charged hadrons per
unit of pseudorapidity obtained at different phase-space kinematic regions ($0<\eta<4.5$) divided by the number 
of wounded constituents, $dN_{\rm ch}/d\eta/Q_{\rm W}$, plotted as function of
centrality expressed via the number of wounded nucleons. The data are from
Refs.~\cite{Adare:2015bua,Bearden:2001qq}.}
\end{center}
\end{figure}

Next, we present our new results for centrality dependence of pseudorapidity density in Au+Au collision at 
RHIC. The BRAHMS~\cite{Bearden:2001qq} and PHENIX~\cite{Adare:2015bua} Collaborations data on $dN_{\rm ch}/d\eta$ 
have been obtained in very different kinematic regions of particle production: from mid-rapidity, $\eta = 0$, up to $\eta =4.5$. 
The number of wounded constituents has been simulated through the use of accordingly modified GLISSANDO 
code~\cite{Broniowski:2007nz,Rybczynski:2013yba}. The obtained linear dependence of $dN_{\rm ch}/d\eta/Q_{\rm W}$ on 
centrality, shown in Fig.~\ref{fig:dndeta}, indicates that the wounded constituent scaling in A+A collisions is universal, i.e., 
independent of the phase-space region in pseudorapidity. 

Our conclusions are as follows:

\begin{itemize}

\item The production of particles at midrapidity follows the wounded constituent scaling, with the quantity
$dN_{\rm ch}/d\eta/Q_{\rm W}$ displaying approximately a flat behavior with centrality. At the LHC the optimal number of
constituents is three.

\item $dN_{\rm ch}/d\eta/Q_{\rm W}$ scales with the number of wounded constituents at different kinematic regions of phase-space.

\item For p+p collisions, the elliptic and triangular eccentricities are large, hence it is expected that the 
corresponding flow coefficients will be large. We also expect that the resulting triangular flow should be
significantly smaller from the elliptic flow.

\end{itemize}

Acknowledgements: Research supported by the Polish National Science Centre grants 2015/17/B/ST2/00101, 2015/19/B/ST2/00937, and 2015/18/M/ST2/00125.


\end{document}